\documentstyle[psfig]{mn}

\newcommand{\eq}{\begin{equation}}
\newcommand{\en}{\end{equation}}
\newcommand{\lab}{\label}

\begin{document}


\title[Pulsar emission]{A Model for `Double Notches' in Radio Pulsar Profiles}

\author[Geoff Wright]{G.A.E.Wright\\
 Astronomy Centre, University of Sussex, Falmer, BN1 9QJ, UK}

\date{Accepted.........Received.....; in original form 2003...}

\maketitle

\begin{abstract}

Aberration and retardation of the radiation within a pulsar's
magnetosphere enable a single corotating absorbing region in the outer
magnetosphere to give rise to a `double notch' in the pulsar's
observed radio profile. This effect requires that a pulsar has not
only absorbing but also emitting regions at heights which are a
significant proportion of the light cylinder radius.  The known
presence of double notches in the weak secondary emission of three
nearby pulsars suggests that this emission comes from open field lines
well above their poles and that an obscuring region extending over no
more than 5 percent of the light cylinder radius is located at or near
the light cylinder.

\end{abstract}

\section{INTRODUCTION}
\subsection{Background}
It is known that at least three pulsars have mysterious double
`notches' in their integrated profiles (McLaughlin $\&$ Rankin 2003,
henceforth MR). All are also known to be sources of pulsed X-rays
(Wang $\&$ Halpern 1997, Zavlin et al 2002), and PSR 1929+10 and PSR
0950+08 also have optical identifications (De Luca et al 2003, Mignani
et al 2002, Zharikov et al 2003). Their high-energy spectra have
variously been interpreted as polar cap thermal emission sustained by
particle backflow or as non-thermal cyclotron emission, with the
latter being a particularly strong indicator of outer magnetospheric
activity. Since one of the three is a millisecond pulsar and the other
two are slow normal pulsars they would seem to have little in common -
even their magnetic fields computed at the light cylinder differ by
several orders of magnitude. The only intrinsic factor all three have
in common is a similar "acceleration parameter", a measure of the
potential difference available to accelerate particles within the
magnetosphere.

The three pulsars are nearby and, possibly because of this, have
detectable weak emission though a large fraction of their periods. The
double notches are embedded in the weak emission and MR report that
the phase of their centroid is frequency-independent, although their
width and separation seems to be frequency-dependent and the feature
may not be seen at all at some frequencies. These are the salient
points on which an explanation must be built.

One piece of obvious and valuable information to be gleaned from the
observations is the physical scale of the phenomenon. This can be
inferred from the notch separation, which MR report to be around
$10^{o}$ for the normal pulsars and $3.3^{o}$ for the millisecond
pulsars. One might seek an explanation in terms of unusual surface
composition or magnetic field structure, which would then imply a
lateral scale very small compared to the polar cap dimensions. But we
would then require that such 'freak' conditions existed on a similar
scale in three very different pulsars. In this paper we offer an
explanation based on the effects of aberration and retardation,
leading to physical scales based on light-travel times. The notches
may thus corespond to lateral or longitudinal features on a scale up
to one sixth of a light cylinder radius (or ~2000km in the case of PSR
0950+08 and PSR 1929+10 and just 20km for PSR J0437-4715), a scale far
greater than inferred height variations for polar cap emission of a
single frequency (Gupta $\&$ Gangadhara 2003, Dyks et al 2003a). This
suggests that the emission in and around the notches comes from
regions higher in the magnetosphere - an idea which is important
since, if true, it would demonstrate support for a long-held suspicion
(e.g. Arons 1979, Hankins $\&$ Cordes 1981, Gil 1985) that not only
new-born but also older normal pulsars are capable of emitting at
radio frequencies in the outer magnetosphere.

To account for notches in such emission it is natural to turn to the
possibility of an obscuring region lying between the emitting region
and the observer. Luo $\&$ Melrose (2001) and Fussell et al (2003)
have suggested that regions of emission suppression can arise in a
pulsar magnetosphere as a result of cyclotron absorption. Clearly any
obscuring feature must lie within the magnetosphere and corotate with
it or the notches would not appear at a fixed phase of the pulse
window. The size of the obscuration estimated from the width of the
individual notches comes out as 600km in the case of PSR 0950+08,
400km in the case of PSR 1929+10 and a tiny 5km for PSR
J0437-4715. Such scales are not impossible to envisage: perhaps a
region of locally-enhanced particle density at high energies located
either at the light cylinder or on the boundary of the corotating zone
defined by the last fieldlines to close within the light cylinder. But
how do we explain $\it{two}$ notches?

It is most unlikely that $\it{two}$ obscuring regions lie adjacent to
one another at any location in the magnetosphere at a fixed or
frequency-dependent separation from one another. It would not
correspond to any known feature from theory or observation. So instead
we turn to one of the unique properties of a pulsar magnetosphere to
provide us with a possible explanation. Pulsar magnetospheres differ
from conventional magnetospheres in that the particles move tightly
bound to magnetic fieldlines and emit their radiation in a
extraordinarily narrow beam.  This implies that most regions of the
magnetosphere are invisible to the observer at any given instant. We
will show in this paper that under these circumstances it is possible
for a single obscuring region to produce a double notch effect, and in
such a way that observed features of the notches enable us to deduce a
simple relation between the location of the emitting and obscuring
regions.

Section 2 summarises the basic principles employed in the model, and
Sections 3 and 4 give the detailed geometric calculations for the
cases of radial and dipolar fieldlines respectively. Section 5 applies
the results to MR's observations of the three candidate pulsars, and
Section 6 summarises the implications and conclusions for future
high-energy observations.
\section{BASIC FEATURES}
\subsection{Model outline}
To demonstrate the central idea of this paper we first consider a
simple two-dimensional model of radially-directed sources moving
relativistically and rotating perpendicularly about a central
axis. The radiation from those lying towards the centre will be
subject to weak aberration, but those further out towards the light
cylinder will have their motion, and hence their radiation, strongly
diverted from the radial vector in the observer's inertial frame. The
further a source is from the centre, the sooner its radiation will be
directed towards the observer, since its greater angular motion will
yield greater aberration. At any given moment, all sources in the
magnetosphere whose radiation is instantaneously directed towards the
observer must lie on a spiral centred on the axis and fixed in
relation to the observer (depicted as a finely-dotted curve in Fig
1). As the magnetosphere turns, sources will pass across this spiral
and, if at an appropriate frequency, will be detected by the observer.

Furthermore, as a consequence of retardation, the sources will be
observed in contrary sequence to which the radial vectors are
presented to the observer (ie counter to the direction of the
rotation) : those most distant from the axis will be seen first, those
near the centre last. In the case of a single radial vector of such
emitters extending almost to the light cylinder, up to a third of the
rotation period will elapse (instead of one sixth if retardation were
neglected) before all of them are seen.  This would be perceived as
nearly $100^{o}$ in phase and would be greater still if the vectors
were not radial but led into the sense of rotation. It is suggested
here that the weak coherent emission observed in a number of radio
pulsars through a wide range of pulse phase arises through this
effect.

Within this two-dimensional picture, we postulate the existence of a
$\it{single}$ corotating site A*, located within the light cylinder
(radius $r_{lc}$) at a distance of $r_{*}$ from the star, which is
capable of absorbing all incident radiation.  Once it subtends less
than a critical angle $\frac{r_{*}}{r_{lc}}$ (thus always less than
one radian) with the observer's line of sight, such a site will
subsequently be able to absorb radiation from sources on the section
of the observer's spiral extending from the rotation axis to the
location of A*, and hence its effect will be detectable by a distant
observer. As it swings towards the observer's line of sight, each
position of A* defines an instant when $\it{two}$ separate sources on
the spiral, one closer to the axis, one closer to A*, emit radiation
which will subsequently be absorbed by A* at the radius
$r_{*}$. Clearly the site closer to the axis will be absorbed later
than the other. In this way A* throws its "shadow" : a line drawn
between the two obscured sites always points at the current location
of A* and always intersects the observer's line of sight at the same
angle ($\frac{r_{*}}{r_{lc}}$). This result is illustrated in Fig 1,
and the formal geometry of the model is set out in Section 3.

\begin{figure}
$$\vbox{
\psfig{figure=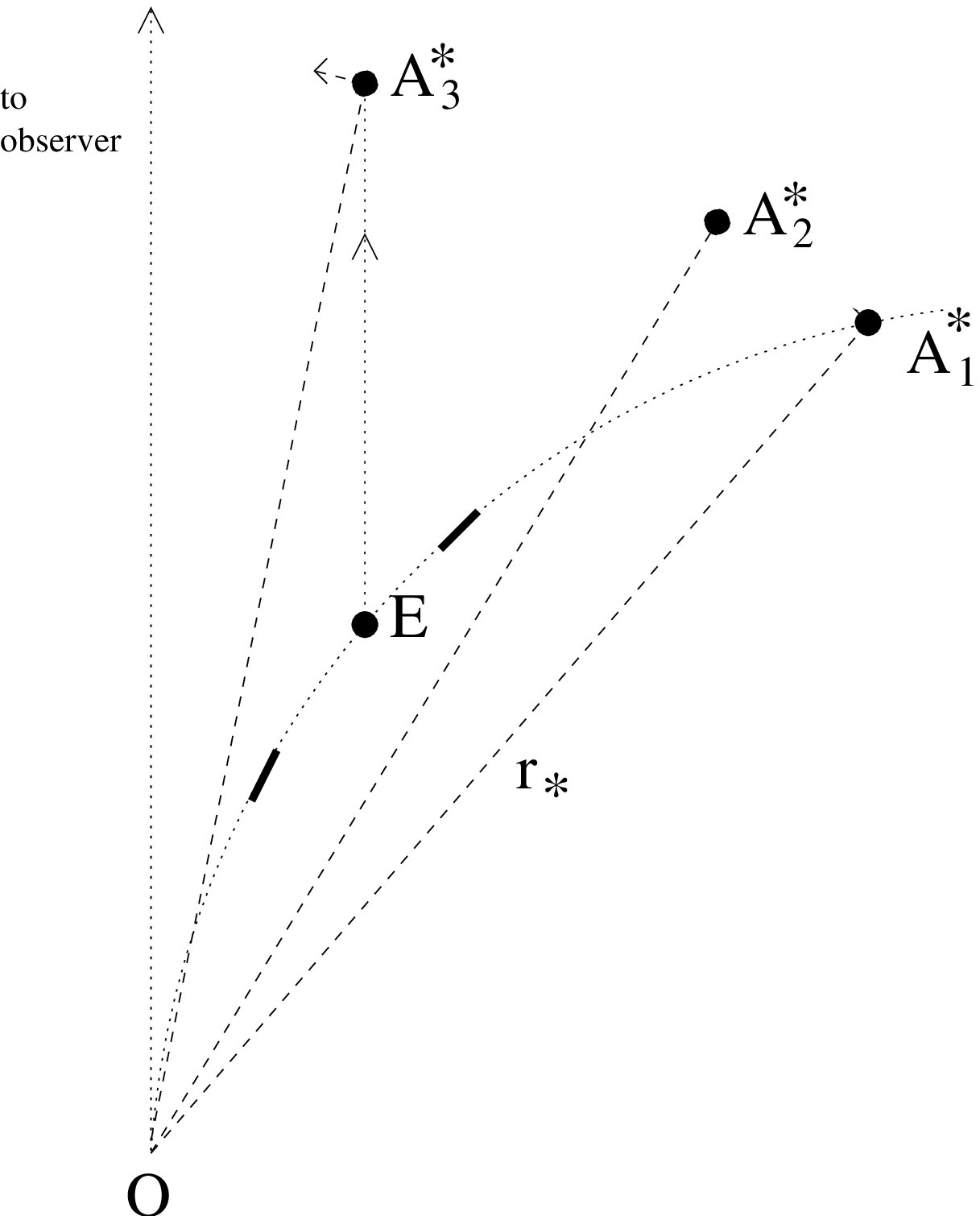,width=6.5truecm,angle=0}
}$$
\caption{The dotted section of a spiral is the fixed locus on which
all sources radiate towards the observer as a result of the effect of
rotational aberration on their relativistic radial motion.
$A^{*}_{1}$ is the initial position of a corotating absorbing site A*
as it crosses the spiral. At any moment when A* is between $A^{*}_{1}$
and $A^{*}_{2}$, sources crossing $\it{two}$ distinct sections
(highlighted) on the spiral will emit radiation in the direction of
the observer to be later absorbed by A*. The changing location of
these sections is defined by the instantaneous intersection of the
spiral and a line parallel to $OA^{*}_{1}$ which passes though the
current position of $A^{*}$. At the instant when A* is at $A^{*}_{2}$
the source at the location E is the final source on the spiral to emit
radiation which can be absorbed (and $EA^{*}$ is tangent to the spiral
and parallel to $OA^{*}_{2}$). The emission from E is absorbed at
$A^{*}_{3}$, where the angle between $A^{*}_{2}$ and $A^{*}_{3}$ is
double that between $A^{*}_{1}$ and $A^{*}_{2}$.}
\label{Fig1}
\end{figure}

Fig 1 summarises the three phases of the process. In the first phase,
as the absorber turns towards the observer's line of sight between
$A^{*}_{1}$ and $A^{*}_{2}$, it defines two changing locations on the
spiral from which the momentary emitted radiation will be later
absorbed. As A* approaches $A^{*}_{2}$ these locations draw together,
finally becoming briefly a single location E (and just before, a
narrow pair), located at a radius $\frac{r_{*}}{2}$, after which no
further emission can take place which will subsequently be absorbed.
At the conclusion of the first phase the absorber will sustain an
angle $\frac{3r_{*}}{4r_{lc}}$ to the line of sight and E will lie
further in, exactly half way between the original angular position of
A* and the observer's line of sight. This crucially determines the
fixed angle between E and A* as $\frac{r_{*}}{4r_{lc}}$.  Compounding
this with the aberration ($\frac{r_{*}}{2r_{lc}}$) and retardation
($\frac{r_{*}}{2r_{lc}}$) accompanying the emission from and around E,
we conclude that the double notches are observed at a phase
$\frac{5r_{*}}{4r_{lc}}$ ahead of the phase of A*.

In phase two, which is exactly twice as long as phase one, A*
continues to rotate and absorb radiation previously emitted from the
upper section of the fixed spiral. When A* is at $A^{*}_{3}$ and
$\frac{r_{*}}{4r_{lc}}$ from the line of sight it absorbs the emission
from E - and just before and after this moment it absorbs in rapid
succession the radiation emitted from the narrow pair of emitters on
either side of E, causing the observer at an appropriate frequency to
see a double notch.

Finally, in a third phase equal in length to the first, A* traverses
the remaining quarter of the arc to the point where it lies precisely
on the observer`s line of sight, and during this time sequentially
absorbs the previously emitted radiation from sources below E at the
midpoint of the spiral down to the surface of the star.

In reality only a small fraction of a pulsar magnetosphere may be
radiating coherently, and certainly not universally at the same
frequency. But the most obvious difference between the above model and
real pulsars is that the fieldlines will not be aligned in a radial
direction. In Section 3 we modify the previous picture so that the
emission which is to be absorbed is directed at the same fixed angle
to the radial vector (although coming from different sources and in
general lying on different fieldlines).  We find that as long as the
angle (either trailing or leading) is no greater than
$\frac{r_{*}}{r_{lc}}$ a double-absorption feature can again be
generated by a corotating absorber. However the absorber will now be
located at a radius greater than the sum of the heights of the
obscured sources for leading fieldlines and less for trailing
fieldlines. Nevertheless it can be shown that the differential between
the observed phase of the notches and the phase of A* is only weakly
dependent on the fieldline inclination.

In the final part of Section 4 it is demonstrated that for emitting
systems with the rotation axis and the observer's line of sight
inclined at an angle $\alpha$, the effects of aberration will be
weaker and the allowed heights of the absorbed emission and the
absorbing region will increase from the perpendicular case by a factor
$\frac{1}{\sin{\alpha}}$. Nevertheless the phase shifts in the pulse
profile can be interpreted using the same formalism.
\subsection{Observational inferences}
In Section 5, armed with the above results, and with $r_{*}$ and
$\alpha$ as the only free parameters, we take the data of the three
pulsars given by MR and attempt to infer the true positions in the
magnetosphere of the emission surrounding the notches and of their
corresponding absorbing regions.

Assuming PSR 1929+10 is a perpendicularly rotating dipole, we
demonstrate that the notch can be generated by an absorbing region
close to the light cylinder and a near-vertical or convex band of
emission ahead of the polar axis, suggesting that the observed weak
inter-pole emission and its position angle sweep may be the
relativistic consequence of such a band or bands.

A similar model was found possible for PSR 0950+08, but without
implying orthogonality.  A single band of emission leading upwards
from the pole (assumed to be located at the main pulse) towards the
light cylinder, may account for the observed weak emission, with the
interpulse representing some feature in the outer magnetosphere near
the light cylinder. The absorbing site would also be at the light
cylinder radius, some $14^{o}$ ahead of the point on the band whose
emission is absorbed. The pulsar should neither be near alignment nor
near orthogonality.

Finally it is shown that the "late" notches of PSR J0437-4715 can be
modelled as long as the pulsar is assumed to be at a low angle of
inclination.  Then both the obscuring and obscured regions must be
relatively close to one another at heights above the star greater than
the light cylinder radius. However, fieldlines are preferred which lie
within the upper open zone of the magnetosphere, but not necessarily
the last open fieldlines.

A summary of the conclusions, dicussing the consequences of these
results for future radio and high-energy observations of these
pulsars, are given in Section 6. It is stressed that the polar cap and
the outer magnetosphere may exist in an interactive state in these and
other pulsars (Wright 2003), as is hinted at by recent simultaneous
multifrequency observations of the Crab (Shearer et al 2003) and Vela
(Donovan et al 2003) pulsars.
\section{Geometric Basis}
To formally demonstrate the geometric principle of double obscuration
we adopt the two-dimensional picture of Fig 2, in which a source S at
a radius r is rotating about O at a fixed angular rate of $\Omega$. In
a rotating frame the emission is radially directed away from O, but in
the inertial frame this results in an aberration component
perpendicular to OS. The resultant direction of the emission is then
instantaneously towards an observer indicated at the top of the page,
making an angle with OS of
\eq
\lab{1}
\phi\approx\sin{\phi}\approx\frac{r\Omega}{c}=\frac{r}{r_{lc}}
\en 
where c is the speed of light and $r_{lc}$ the light cylinder radius.

A site $A^{*}$ at which all incident radiation is assumed to be
absorbed is also rotating at a rate $\Omega$ and located at a distance
$r_{*}$ and making a permanent angle of $\Theta$ with OS. If at a time
t=0 S is positioned so that its radiation (subject to aberration) is
directed towards the observer, then the radius vector of the absorber
$A^{*}$ will at that moment make an angle $\theta$ with the direction
to the observer, given by
\eq
\lab{2}
\theta=\Theta+\phi\approx\Theta+\frac{r}{r_{lc}}
\en 
Since $(\theta-\Theta)$ is the angle the source's postion vector makes
with the observer's line of sight, (2) defines the equation of the
critical spiral.  We need to know for which coordinates (r,$\Theta$)
of S will the radiation from S be absorbed before it reaches the
observer (ie we seek to know from which position or positions on the
spiral will the radiation be absorbed by A*). Represented in Fig 2, this
requires that $A^{*}$ has rotated to a new position $A^{'}$ in the
same time that the radiation has travelled from S to $A^{'}$. If
$\phi{`}$, the angle between the new vector $OA{'}$ and the t=0 vector
for S, is assumed to be small (as it will be if the aberration effect
is small), then the distance travelled by the radiation before
absorption will be ($r_{*}-r$) and $\phi{'}$ is given in the small
angle approximation by
\eq
\lab{3}
\frac{\phi{'}}{(r_{*}-r)}=\frac{\phi}{r_{*}}
\en 
hence from (1)
\eq
\lab{4}
\phi{'}=\frac{(r_{*}-r)r}{r_{lc}r_{*}}
\en 

The crucial condition that A* rotates to $A{'}$ in the same time, T, 
that the radiation travels from S to $A{'}$ is then
\eq
\lab{5}
T=\frac{\Theta+\phi{'}}{\Omega}=\frac{r_{*}-r}{c}
\en 
which yields from (4)
\eq
\lab{6}
\Theta=\frac{(r_{*}-r)^2}{r_{lc}r_{*}}
\en 
and, substituting from (2), we obtain a quadratic for r:
\eq
\lab{7}
r^2-r_{*}r+r_{*}(r_{*}-\theta r_{lc})=0
\en 
\subsection{Double Obscuration}
The two solutions $r_{1},r_{2}$ of equation(7) represent two possible 
locations at which an emitter S can be obscured by $A^{*}$ as it rotates.
These solutions satisfy
\eq
\lab{8}
r_{1}+r_{2}=r_{*}
\en
and for both these roots to be positive we additionally require that 
$r_{1}r_{2}>0$, or
\eq
\lab{9}
\theta<\theta_{max}\approx\frac{r_{*}}{r_{lc}}
\en
Given that $\theta_{max}$ must be less than unity, we have our first
constraint on the position of the absorber, namely that if radiation
is to be absorbed then it must be emitted when the absorber's radial
vector bears an angle less than one radian (about $57^{o}$) to the
observer's line of sight, and in practice much less.
 
The difference (i.e.the approximate distance) between the two locations is
\eq
\lab{10}
r_{1}-r_{2}=\epsilon=r_{*}(4\frac{\theta r_{lc}}{r_{*}}-3)^{\frac{1}{2}}
\en 
so that
\eq
\lab{11}
r_{1,2}=\frac{r_{*}\pm\epsilon}{2}
\en
and the physical condition that $\epsilon$ is real, together with (9) 
above, establish the upper and lower limits on $\theta$ as
\eq
\lab{12}
\theta_{min}<\theta<\theta_{max}
\en

\begin{figure}
$$\vbox{
\psfig{figure=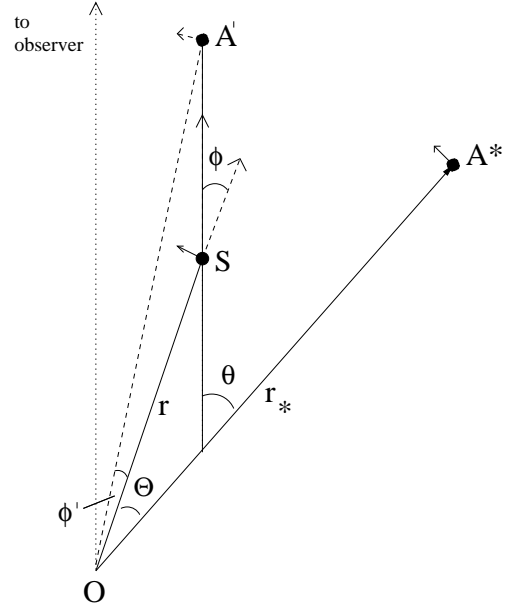,width=6.5truecm,angle=0}
}$$
\caption{The source S, whose motion is a combination of a 
relativistic radial component and a rigid rotation about A, is at an
instantaneous angle $\Theta$ to an absorber A* rotating
perpendicularly about an axis through O. The emission from S is
intercepted by A* at $A^{'}$.}
\label{Fig2}
\end{figure}

where
\eq
\lab{13}
\theta_{min}\approx\frac{3r_{*}}{4r_{lc}}
\en
ie only when A* is between these two angular positions can radiation
be absorbed. This defines the duration of Phase 1, at the conclusion
of which $\epsilon\rightarrow{0}$ and the two possible solutions for
the position of S coalesce to the position E in Fig 1.

The upper limit is only appropriate when $r_{1}\rightarrow{r_{*}}$ and
$r_{2}\rightarrow{0}$. The more interesting case here is when $r_{1}$
and $r_{2}$ are close to each other and $A^{*}$ is located at
$\theta_{min}$. Through Phase 1, as $\theta$ decreases from
$\theta_{max}$, the positions of $r_{1}$ and $r_{2}$ gradually move
along the spiral towards each other, meeting at the lower limit of
(13) and fixing the location of E in Fig 1 on the critical
spiral. Hence in general a narrow extended absorbing region centred
radially or azimuthally on A* will at any moment obscure two strips of
the dotted spiral on either side of the midpoint, as indicated in Fig
1.
\subsection{The relative geometry of the two obscured regions}
The location of $r_{1}$ is at an angle 
\eq
\lab{14}
\Theta_{1}=\frac{(r_{*}-r_{1})^2}{r_{lc}r_{*}}=\frac{r_{2}^2}{r_{lc}r_{*}}
\en
to the vector of A* (and $r_{1}$ is at the equivalent angle
$\Theta_{2}=\frac{r_{1}^2}{r_{lc}r_{*}}$).  Thus at the moment of
emission the angle $\Theta_{12}$ between $r_{*}$ and the midpoint of
$r_{1}$ and $r_{2}$ is
\eq
\lab{15}
\Theta_{12}=\frac{r_{1}^2+r_{2}^2}{2r_{lc}r_{*}}
\en
so that at or close to the conclusion of phase 1, when 
$r_{1}\approx{r_{2}}\approx{\frac{r_{*}}{2}}$, we can establish the angle
between the observed mid-point of the notches and the site of the absorber as
\eq
\lab{16}
\Theta_{12}\approx\frac{r_{*}}{4r_{lc}}=\frac{\theta_{min}}{3}
\en
Thus $\Theta_{12}$ is less than $15^{o}$, and 
$\theta_{min}$ can never be more than $45^{o}$.

To measure the duration of phase 2 we need to evaluate $\phi^{'}$, the angle 
between the initial position of the emitter and the position at which its 
emission is absorbed. It is the same for both emitters:
\eq
\lab{17}
\phi^{'}=\frac{(r_{*}-r_{1})r_{1}}{r_{lc}r_{*}}\approx
\frac{r_{1}r_{2}}{r_{lc}r_{*}}
\en
When $r_{1}$ and $r_{2}$ are close to the midpoint, both $\phi^{'}$
and $\Theta_{12}\approx\frac{r_{*}}{4r_{lc}}$. So in geometric terms
OS bisects the angle between $OA^{'}$ and OA*, and in phase 2 A*
rotates through $\phi^{'}+\Theta_{12}\approx\frac{r_{*}}{2r_{lc}}$,
twice the angular duration of phase 1.

We can show that the line joining the two solutions always points at
the current position of A* and makes an invarying angle to the
observer's line of sight. The angle SA*O which S makes to the
absorber's vector is the same for both solutions since it is given by
\eq
\lab{18}
\frac{r_{1}\Theta_{1}}{r_{2}}= 
\frac{r_{2}\Theta_{2}}{r_{1}}=\frac{r_{1}{r_{2}}}{r_{lc}r_{*}}
\en
(from (14)), and the direction SA* makes with the observer's line of sight 
($A^{'}SA^{*}$) is, from (2) and (18),
\eq
\lab{19}
\theta+\frac{r_{1}r_{2}}{r_{lc}r_{*}}=\frac{r_{*}}{r_{lc}}
\en
which is the same for all $r_{1},r_{2}$, and implies that the two
obscured sources always point towards A* and remain aligned parallel
to the original direction of A*. The line between the sources acts as
a kind of "shadow" of A* which passes across the observer's
spiral. But the "shadow" moves rapidly, passing over the entire spiral
during phase 1, as A* physically rotates through an angle
$\frac{r_{*}}{4r_{lc}}$. By then it has crossed fieldlines from the
original position of OA* ($OA^{*}_{1}$ in Fig 1) to OS, which is now
an angle $\frac{r_{*}}{2r_{lc}}$ (OE in Fig 1) to the observer's line
of sight. Hence the "shadow" can be perceived as moving at twice the
rotating speed of the fieldlines, continuing to overtake them until it
passes through E (see Figs 1 and 3).

This has the further consequence that an absorber extended along or
perpendicular to OA* will not have its "shadow" on the observer's
spiral expanded by the geometry, but will project its size along
parallel lines. Hence the observed notch width can be taken as an
approximate measure of the absorber's dimensions.

Phase 2 closes with the emission from E absorbed by A*, which is now
at $A_{3}$ in terms of Fig 1. In Phase 3, the absorber moves to
position itself precisely between the observer and the star, and as it
turns absorbs the radiation earlier emitted from the lower section of
the fixed spiral, finally absorbing the emission from O.

We are now able to evaluate the crucial time intervals necessary for
interpreting the observed apparent phase of the double notches in a
pulsar profile. First we need the phase delay $\psi_{em}$ between the
emission from E and the moment at which it absorbed at $A^{*}_3$
(i.e. the end of Phase 2). This is the sum of the emission's
retardation ($\frac{r_{*}}{2r_{lc}}$) and its aberration
($\frac{r_{*}}{2r_{lc}}$), giving
\eq
\lab{20}
\psi_{em}=\frac{r_{*}}{r_{lc}}
\en 
Then from (16) we know that the fixed angle between the midpoint of
the notches and the alignment of the absorber is
$\frac{r_{*}}{4r_{lc}}$, so we can finally say that in an aligned
rotator of radially beamed emitters any observed double notches will
lead the true position of their absorbing site by a phase of
\eq
\lab{21}
\psi_{abs}=\frac{5r_{*}}{4r_{lc}}
\en
Both (20 and (21) are only accurate if $r_{*}\ll{r_{lc}}$, and can
only be approximate in our later attempts to consider cases where
$r_{*}\rightarrow{r_{lc}}$.

\section{Modifications for dipole geometry}
Before we can apply the above results to the three pulsars described
by MR in the companion paper, we first consider how they must be
modified to describe sources which are constrained to radiate parallel
to dipolar field lines, and furthermore that the dipole may take any
angle to the rotation axis. We consider each of these points in turn.
\subsection{Non-radial fieldlines} 
Firstly, although dipolar fieldlines will be non-radial and splay
apart from the polar cap, we will continue to assume that the observed
fieldlines lie in a two-dimensional plane (as in the case of a
perpendicular rotator).  But in general we will assume that at every
position the source radiates at an angle $\Phi$ ($\ll{1}$) to the
radial direction. The sign of $\Phi$ indicates whether we are
approximating trailing ($>0$) or leading ($<0$) fieldlines, and $\Phi$
will in general be both radially and azimuthally dependent, and hence
time-dependent.

To direct radiation from the source S in Fig 2 towards the observer,
aberration must now shift the propagation vector not through $\phi$
but through the total of $\phi$ and $\Phi$, which may be greater
(trailing) or less (leading) than $\phi$. This modifies our result (1)
to
\eq
\lab{22}
\phi+\Phi\approx\sin(\phi+\Phi)\approx\frac{r\Omega}{c}=\frac{r}{r_{lc}}
\en
so that now the deviation due to aberration is
\eq
\lab{23}
\phi=\theta-\Theta=\frac{r}{r_{lc}}-\Phi
\en
Note that if $\Phi<\frac{r}{r_{lc}}$ the aberration is still ahead of
the radial vector. But for $\Phi>\frac{r}{r_{lc}}$ the sources do not
emit in the direction of the observer until their radial vectors have
passed through the observer's line of sight. If $\Phi$ is constant,
(23) is the equation of a new fixed spiral, analogous to that in Fig
1, shifted by the fixed angle in the sense of the rotation if
$\Phi<0$, and against it if $\Phi>0$.

Following the argument of the previous section, we again apply the
condition that the absorber rotates precisely to meet the
observer-directed radiation, and this yields a modified value for the
fixed angle between OA* and OS:
\eq
\lab{24}
\Theta=\frac{(r_{*}-r)^2}{r_{lc}r_{*}}+\frac{(r_{*}-r)}{r_{*}}\Phi
\en 
whence we obtain a new quadratic for r:
\eq
\lab{25}
r^2-(r_{*}+\Phi{r_{lc}})r+r_{*}(r_{*}-\theta r_{lc})=0
\en 
So far we have allowed $\Phi$ to be a radially and azimuthally
dependent function.  This would make (25) a model-dependent and
time-dependent equation to yield the locations of the sources whose
emission is absorbed.  However, since for observational reasons we are
ultimately interested in solutions for small emission regions close to
E over which the magnetic field direction varies little, we will now
take $\Phi$ as a constant. The two roots of (25) are related by
\eq
\lab{26}
r_{1}+r_{2}=r_{*}+\Phi{r_{lc}}
\en
and
\eq
\lab{27}
r_{1}r_{2}=r_{*}(r_{*}-\theta{r_{lc}})
\en
We note immediately from (26) that for absorption to occur at two
sites we mqust place the important constraint between $\Phi$ and the
radius of the absorber that
\eq
\lab{28}
|\Phi|<\frac{r_{*}}{r_{lc}}
\en
so that neither $r_{1}$ nor $r_{2}$ shall exceed $r_{*}$ (for
$\Phi>0$) nor be negative (for $\Phi<0$). In what follows, we will
therefore find that the sign-dependent ratio defined by
\eq
\lab{29}
k=\Phi\frac{r_{lc}}{r_{*}}
\en
with $k<1$, will be useful in highlighting the differences between the 
cases of radial and non-radial fieldlines.

However there is one important feature of the radial solution which
remains unchanged. If we evaluate $\Theta_{1}$ from (24) and (26) we
now obtain
\eq
\lab{30}
\Theta_{1}=\frac{(r_{*}-r_{1})r_{2}}{r_{lc}r_{*}}
\en
from which, using (23), we again have the angle S obtains at A* 
\eq
\lab{31}
\frac{r_{1}\Theta_{1}}{(r_{*}-r_{1})}=\frac{r_{1}{r_{2}}}{r_{lc}r_{*}}
\en
which is the identical result to (18) and is the same for both obscured 
sources. And once more the line between the sources and the absorber 
makes an angle of $\frac{r_{*}}{r_{lc}}$ with the observer's line of sight. 

\subsection{Modified Phases}
Although the observer's spiral will be shifted by an angle $\Phi$
either counter (leading fieldlines) or with (trailing fieldlines) the
direction of rotation, it is clear from equation (26) that in either
case only for $\theta<\theta_{max}=\frac{r_{*}}{r_{lc}}$ can there be
two physical roots to equation (25). As $\theta$ reduces, the double
roots will begin with those at $r_{1}=r_{*}+\Phi{r_{lc}}$($<r_{*}$)
and $r_{2}=0$ and end with
\eq
\lab{32}
r_{1,2}=\frac{r_{*}}{2}(1+k)\pm\frac{\epsilon}{2}
\en
in the limit $\epsilon\rightarrow{0}$. This is the modified position
of E, the limiting location of S on the spiral and the midpoint of the
notches. Note how if it is located on leading fieldlines it will be
closer to the surface of the star (for a given $r_{*}$) and further
from it for trailing fieldlines. Then (by substitution in (25)) the
absorber will be at the angle to the line of sight of
\eq
\lab{33}
\theta_{min}=\frac{r_{*}}{r_{lc}}(\frac{3}{4}-\frac{k}{2}(1+\frac{k}{2}))
\en
so that the total change in $\theta$ in Phase 1 is now
\eq
\lab{34}
\theta_{max}-\theta_{min}=\frac{r_{*}}{r_{lc}}(\frac{1}{4}+\frac{k}{2}
(1+\frac{k}{2}))
\en
Given the constraint (28), even when $k<0$ (leading fieldlines) this
quantity will never be less than zero. However it may be greatly
reduced, limiting the time when double absorption is
possible. Conversely, trailing fieldlines extend this possibility.

Setting $r_{1}=r_{2}=\frac{r_{*}(1+k)}{2}$ into (30) we see that the
source at E in Fig 1 will now lie at a fixed angle ahead of the
absorber by
\eq
\lab{35}
\Theta_{12}=\frac{r_{*}}{4r_{lc}}(1-k^2)
\en
reduced in size and independent of the sign of $\Phi$ on the result (16). 

The time travelled by radiation from the modified position of E, given
by the limit value of (32), to the absorber yields a modificaton in
the angular duration of Phase 2:
\eq
\lab{36}
\frac{r_{*}-\frac{1}{2}r_{*}(1+k)}{r_{lc}}=\frac{r_{*}}{2r_{lc}}(1-k)
\en
In the case of leading fieldlines this is a significant increase on
the equivalent angle in the purely radial case, because the reduction
in the height of E means the time needed for the radiation to reach
the absorber is longer. The effects are exactly opposite for trailing
fieldlines.

We are again in a position to calculate the apparent phase shift
between the observed location of the double notches and the "true"
phase in the magnetosphere of the vector defining their
midpoint. Through aberration the midpoint is observed
$\frac{r_{*}}{2r_{lc}}(1-k)$ too soon (obtained by setting
$r=\frac{r_{*}}{2r_{lc}}(1+k)$ in (23)), and through retardation a
further $\frac{r_{*}}{2r_{lc}}(1+k)$, giving a total of
$\psi_{em}=\frac{r_{*}}{r_{lc}}$, identical to equation (20).  Note
that this quantity is independent of k for a given $r_{*}$: a
reduction in aberration is compensated by an increase in retardation
and $\it{vice versa}$.

The additional phase separation between the midpoint's vector and the
position of the absorber is $\frac{r_{*}}{4r_{lc}}(1-k^2)$ from
equation (35), giving a total shift of
$\psi_{abs}=\frac{5r_{*}}{4r_{lc}}(1-\frac{k^2}{5})$, replacing
equation (21). Thus the modification of this value compared with the
radial case is small and independent of the sign of k.

\subsection{Non-orthogonal rotators}
In our deliberations so far the absorber, the sources and the
observer's line of sight all lie in the same plane orthogonal to the
rotation axis. If our line of sight is not orthogonal to the rotation
axis then an absorber rigidly circling the axis will only obscure a
source whose emission, following aberration, lies in the
near-instantaneous plane defined by the absorber and the line of
sight. For this then to create a $\it{double}$ absorption feature is
probably impossible in the poloidal fieldlines of a perfect dipole,
but it cannot be excluded in the real field configuration containing
swept-back poloidal components.

A full consideration of the 3-dimensional geometry would make the
analysis extremely complex. However the closeness of the notches means
that to a reasonable approximation we can again consider a geometry
similar to Figs 1 and 2, where now all directions lie in approximately
the same plane tangential to the cone formed by the rotating absorber
with the star at its apex.
 
Assuming our line of sight passes close to the pole of the pulsar
whose dipole axis is inclined at a general angle $\alpha$ to the
rotational axis, the aberration angle (1) at a given radial distance
from the star must be modified by the factor $\sin{\alpha}$, giving
\eq
\lab{37}
\phi\approx\frac{r\Omega\sin{\alpha}}{c}=\frac{r}{r_{lc}}\sin{\alpha}
\en 
In the plane of our geometry the replacement of $\Omega$ with
$\Omega\sin{\alpha}$ will effectively extend the corotating limit to
$\frac{r_{lc}}{\sin{\alpha}}$ and enable both the absorber and the
obscured sources to be at greater distances from the star.  The
quadratic (7) giving the two heights of the absorbed sites will only
be modified in its final term, giving
\eq
\lab{38}
r^2-r_{*}r+r_{*}(r_{*}-\frac{\theta r_{lc}}{\sin{\alpha}})=0
\en
so the sum of its roots will still be $r_{*}$ and hence the midpoint
of the notches will still be at about half the distance of the
absorbing site. The three phases of Fig 1 will occur in the same
duration since although all angles in Fig 2 are reduced by the factor
$\sin{\alpha}$ they are rotated more slowly with respect to the
observer. This implies also that the corrections for the true phases
in the pulse window of the observed notch sites (20) and the absorbing
region itself (21) remain unchanged at $\frac{r_{*}}{r_{lc}}$ and
$\frac{5r_{*}}{4r_{lc}}$. However $r_{*}$ can now be up to
$\frac{r_{lc}}{\sin{\alpha}}$ allowing the observed phase shifts in
pulsars closer to alignment to greatly exceed the formal limits of 1
and $\frac{5}{4}$ radians.
\section{Applications to pulsars} 
\subsection{Notch formation}
In real pulsars the beamed emission at a particular frequency will lie
on a strip stretched across many fieldlines. The observer will
perceive this radiation as the strip crosses his 'fixed' spiral,
itself determined by the changing inclination of the dipolar
fieldlines presented to him. At the magnetic poles close to the
surface the fieldlines are near-radial and the emission strips roughly
perpendicular to them, based on a radius-to frequency mapping
(although aberration and retardation effects on the pulsar profile
make identifying the correct mapping difficult to judge (Dyks et al
2003a). In the outer magnetosphere we are dealing with diverging
non-radial fieldlines (to first order those of a rigidly rotating
dipole) and emission regions which in general possess both radial and
azimuthal structure. Fig 3 shows an emitting region crossing the
observer's "spiral" and the presence of an absorbing site which casts
a rapidly-moving "shadow" over the rotating strip of emission, thereby
causing the observer to see a double notch.

Note that an immediate implication of Fig 3 is that the (black)
emission region at the observing frequency must be both thin (ie no
thicker than the passing shadow) and quasi-radial in structure for the
double notch to be formed without blurring.

\begin{figure}
$$\vbox{
\psfig{figure=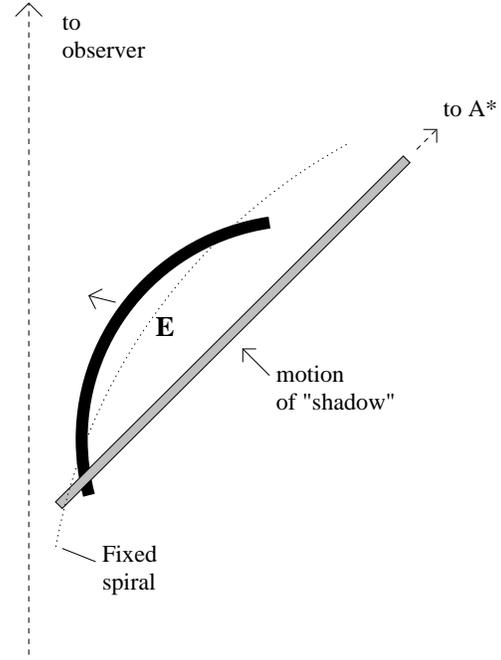,width=6.5truecm,angle=0}
}$$
\caption{The formation of a double notch. The dark band represents 
emission region of a particular radio frequency which is emitted
towards the observer as it rotates across the fixed (dotted line)
spiral. The grey band defines the "shadow" of the absorbing site
A*. Emission from the points where it intersects the spiral will be
later absorbed. The shadow does not rotate, but moves linearly and
rapidly across both the spiral and the emission band, always
maintaining a fixed angle to the observer's line of sight as A*
rotates.}
\label{Fig3}
\end{figure}

\subsection{PSR 1929+10}
We first consider PSR 1929+10 because it is the best candidate for a
perpendicular rotator, and because its double notch feature is part of
a single smooth stretch of emission, which can guide us in our
interpretation. Its poles are nearly $180^{o}$ apart at all
wavelengths (Hankins and Fowler 1986, Rankin and Rathnasree 1997) and
the pulsar has been observed to emit pulsed thermal X-ray emission,
which is difficult to reconcile with an aligned configuration (Wang
and Halpern 1997). However the efforts of numerous authors, encouraged
by the smooth swing of the position angle in the weak extended radio
emission, to fix the pulsar's angle of inclination and our line of
sight direction, have met with conflicting results (Naranyan \&
Vivekanand 1982, Lyne \& Manchester 1988, Blaskiewicz et al 1991,
Rankin 1993a,b, Everett $\&$ Weisberg 2001). We attempt to resolve
this conflict by suggesting that the weak extended emission (Perry \&
Lyne 1985) is not part of the low-altitude polar cap emission and its
single rotating vector system (Radhakrishnan $\&$ Cooke, 1969) but
originates at higher altitudes and requires a separate system fit
which takes relativistic effects into account.

The $10^{o}$ angular separation of the notches suggests in our present
model that the regions of obscured emission are 2000km apart (a sixth
of the light-cylinder radius) - far in excess of variations in
emission heights at a single frequency in conventional polar cap
models (eg Sturrock 1971).  We must therefore look for the site of the
absorbed radiation in the outer magnetosphere, possibly located on the
open field lines emanating from one of the poles, bearing in mind
that, since the absorber itself cannot lie outside the light cylinder,
the absorbed source can be no more than halfway to the light cylinder.

Aberration and retardation mean that the notches feature appears in
the profile well ahead of the "true" phase, $\psi$, of its radius
vector. However, adopting an orthogonal geometry, we can use the pulse
and interpulse to mark the correct phases of the magnetic
poles. Noting that the $\it{interpulse}$ trails the notches by
$85^{o}\approx{1.5} $ radians, we may assume that it is on the open
fieldlines of this pole that the source is located. Then equations
(20) and (21) can be used to constrain $\psi$ and $r_{*}$. If $\psi$
is measured negatively ahead of the pole, and is zero at the pole,
then from (20) (which from Section 3.2 is valid whatever the fieldline
orientation at the emission site)
\eq
\lab{39}
 \psi=\psi_{em}-1.5=\frac{r_{*}}{r_{lc}}-1.5
\en
The minimum value of $|\psi|$, fixed by allowing
$r_{*}\rightarrow{r_{lc}}$, is therefore $\approx{0.5 radians}=28^{o}$
ahead of the pole, at a radius of half the light cylinder, with an
absorbing region at the light cylinder at phase
$\psi+\frac{1}{4}=-14^{o}$. The maximum value of $|\psi|$ can be
determined by constraining the emission to lie on an open fieldline:
if the notch emission is located on the last closed fieldline then
\eq
\lab{40}
\frac{\sin^2{\psi}}{r_{1,2}}=\frac{s^2}{r_{lc}}
\en
where $r_{1,2}$ is the limiting value of equation (32) and s, a
measure of the fieldline's distance from the polar axis, is unity. In
dipole geometry the angle the leading fieldlines make with the radial
vector is $\Phi=\frac{\psi}{2}$, so we have from (32) and (29)
\eq
\lab{41}
r_{1,2}=\frac{r_{*}}{2}(1+\frac{\psi{r_{lc}}}{2r_{*}})
\en
Solving (39), (40) and (41) gives $\psi\approx{-34^{o}}$ and
$r_{*}\approx{0.9r_{lc}}$. Thus now the emitting region is now $6^{o}$
further from the pole but (from (40)) at a somewhat reduced radius of
$0.3r_{lc}$, and the absorbing site is still close to the
light-cylinder, leading the pole by $20^{o}$.

\begin{figure}
$$\vbox{
\psfig{figure=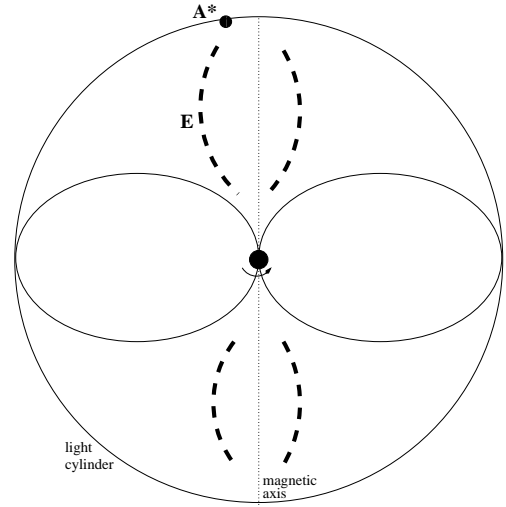,width=6.5truecm,angle=0}
}$$
\caption{A possible configuration for 1929+10. Assuming this pulsar 
to be an orthogonal rotator, and that the absorbing site A* lies near
the light cylinder, then A* will precede the projected magnetic polar
axis and the location E of the absorbed emission will be $15^{o}$
earlier at just under half the light cylinder radius. This suggests
that the weak "pedestal" radio emission from this pulsar may come from
quasi-radial bands on either side of the polar axis, stretching to
near the light cylinder.}
\label{Fig4}
\end{figure}

Whichever of these not dissimilar extremes is correct, we have to
conclude for this pulsar that we are indeed dealing with emission
regions well away from the polar cap, possibly linked to the observed
high-energy emission. The observed emission band in which the notches
are embedded must cross the observer's spiral in a roughly convex form
to facilitate the formation of the notch (Fig 3), and extend in a
narrow vertical strip (Fig 4) so as to give rise to the observed
smooth and slow variation in the position angle. How this is then
precisely linked to the polarization properties of the full $360^{o}$
emission features will be considered in a further paper. Work by Dyks
et al (2003b) shows that emission strictly held to the last closed
fieldlines will create position angle swings more complex than those
observed, so it would seem that a configuration more like the
quasi-radial cigar-shaped emission regions of Fig 4 should be
preferred, in keeping with the picture of a relativistic caustic wave
described in the opening discussion of this paper. Only numerical
modelling (eg Dyks $\&$ Rudak 2003, Dyks et al 2003b) can resolve this
point.

The absorbing region near the light-cylinder can have a size no
greater than the width of an individual notch, which is
$\approx{4^{o}}$ or equivalently $.01r_{lc}$ or 120km.  Since it leads
the polar axis it may be identified as a location where particles in a
polar beam acquire high Lorentz factors in the rotating frame and
leave fieldlines to avoid becoming superluminal in the inertial frame.

\subsection{PSR 0950+08}
The orientation of this pulsar has been subject to much debate over
the years and is far from understood (Hankins $\&$ Cordes 1981, Gil
1983, Lyne $\&$ Manchester 1988, Rankin 1993a,b, von Hoensbroech $\&$
Xilouris 1997, Everett $\&$ Weisberg 2001). In contrast to PSR
1929+10, its two principal radio emission components are separated by
much less than $180^{o}$ and it is difficult to argue that they are
different poles. However this separation does not vary with frequency
(Hankins $\&$ Fowler 1986) as would be expected of a wide single cone
close to alignment, and an intriguing correlation has been found
between pulses in the the main component and the $\it{subsequent}$
pulse in the interpulse component (Hankins $\&$ Cordes 1981). Like PSR
1929+10, a double peak periodicity has been detected in the X-ray
emission and in our interpretation here we will assume that it is not
near alignment and that the main pulse is at the phase of a magnetic
pole.

This pulsar has in common with PSR 1929+10 that it possesses a band of
weak emission extending between the interpulse and main
pulse. Although at first glance the double notch feature - located on
the leading edge of the main pulse - seems not part of this band, on
closer inspection it has been demonstrated to arise in the weak pulse
system which generates this band (reported by Nowakowski et al 2003 -
see MR).  The angular separation of the notches is again about
$10^{o}$, and since the two pulsars have very similar periods this
corresponds to a similar physical separation of 1800km. They also have
similar individual widths, so the size of the absorber will be again
around 400km. We thus interpret the weak bridge of emission, and its
continuation in the main pulse and interpulse, as being generated in
the upper magnetosphere, and we again suggest that in modelling the
position angle swing the bridge and main pulse emission should be
considered as belonging to separate, if interlinked, systems.

In considering this pulsar we adopt the same approach as with PSR
1929+10.  The notches appear just $30^{o}\approx{0.5}$ radians before
the main pulse, but that does not imply that they originate on
leading-edge fieldlines corresponding to the presumed pole at the main
pulse phase, since retardation and aberration may have shifted them to
earlier phase by considerably more than $30^{o}$. Recalling from
Section 3.3 that observed phase shifts are formally independent of the
unknown $\sin{\alpha}$, and replacing 1.5 in equation (39) with 0.5 we
may attempt to see if the notches could arise on the pole's last
closed fieldlines. However the left hand side of (39) has to include
the factor $\frac{1}{\sin{\alpha}}$ to express the mapping of the
dipole coordinate onto the pulse phase:
\eq
\lab{42}
\frac{\psi}{\sin{\alpha}}=\frac{r_{*}}{r_{lc}}-0.5
\en
Equation (40) remains unchanged, but (41) must also reflect the mapping of $\psi$: 
\eq
\lab{43}
r_{1,2}=\frac{r_{*}}{2}(1+\frac{\psi{r_{lc}}}{2\sin{\alpha}{r_{*}}})
\en
We can attempt to solve (42), (43) and (40) to see if the notches can
possible lie on the last closed fieldlines (s=1), allowing for
solutions on both the leading and trailing sides of the pole, but we
find that for all values of $\alpha$ the leading side solution is
unphysical since the height of the emission would be at most 0.06 of
the light cylinder radius, less than the separation of the notches,
and there is no solution at all on the trailing side (ie retardation
and aberration on that side could never conspire to shift the emission
to its observed phase). However, physically plausible solutions are
possible if the emission lies on the quasi-radial fieldlines
surrounding the pole (so that the effects of curved fieldlines are
minimised and $k\ll{1}$). For example, if we assume that the notches
arise from emission on the polar axis (s=0) then $\psi=0$ in (42) and
$r_{*}\approx{0.5r_{lc}}$, so that the emission is at one quarter of
the light cylinder radius. Again, this result is independent of
$\alpha$. The polar region within which solutions are possible is
gradually reduced to a smaller and smaller bundle of (trailing)
fieldlines ($s<1$) as the inclination lessens, and the location of the
supposed absorbing region is shifted to $45^{o}$ or more, trailing the
pole at a distance from the star several times the light cylinder
radius.

Thus we may conclude that the emission obscured in the notches is
coming from within a few degrees of the polar axis, and at a height
well above the star surface. Within this constraint there are many
possible dipole alignments and locations of the absorbing site which
can mimic the observations.  Establishing the correct configuration
requires detailed modelling and comparison with polarisation data, but
the suggestion of this model, supported by the very distinctive
emission stretching for some $150^{o}$ before the main pulse, is that
we are dealing with a radially-extended band of weak emission
stretching from the pole towards the light cylinder, possibly ending
at the location of the absorbing region. If so, the observed
interpulse must define the upper limit of the band, being the closest
point in the magnetosphere to the observer at the moment of emission,
and the bridge of emission maps the band down to the pole. This
configuration has the advantage of naturally explaining the delayed
main pulse to interpulse correlation (Hankins \& Cordes 1981) - since
the interpulse emission is the upper magnetosphere image of the polar
activity - but may require $\sin{\alpha}$ to be substantially less
than 1 - possibly around $\frac{1}{2}$, equivalent to an inclination
of $30^{o}$, to yield the observed long stretch of emission.

\subsection{PSR J0437-4715}
The notches observed in this pulsar are the most difficult to
interpret because so little is known about the orientation and
structure of its magnetic field. Its X-ray detection (Zavlin et al
2002) shows a single-hump time curve and is suggestive of a
non-dipolar structure, and this picture has theoretical support from
models of evolving magnetic fields in old neutron stars (Chen et al
1998), which propose a nearly aligned but "squeezed" magnetic
axis. Gil $\&$ Krawczyk (1997) have, not dissimilarly, modelled the
radio emission with an dipole inclined at $20^{o}$, with only one pole
visible to us.  However the observed polarisation is complex and gives
no simple interpretation in terms of the rotating vector model
(Radhakrishnan $\&$ Cooke 1969, Navarro et al 1997).

To explain the notches noted in MR without abandoning our absorption
hypothesis we might be tempted to extend our intra-light cylinder
assumption. Intriguingly the profile of Fig 2 in MR reveals a number
of single notches in addition to the double notch feature. Thus we
could allow the absorbing agent to be the star itself and assume that
the radiation is emitted by downflowing particles in the back
magnetosphere. Alternatively we may contrive absorbing regions outside
the light cylinder which corotate superluminally through which
particles flow counter to the rotation.Without discarding these
possibilities, we will nevertheless demonstrate below a
self-consistent geometry which can still produce the double notches
within the framework of our basic model.

The positioning of the notches in this pulsar is late ($70^{o}$). So
if we wish the location of the obscured sources to be on open
fieldlines above the visible pole then it is necessary to assume near
alignment in order to interpret the observed phase as a greatly
expanded projection of the underlying dipole phase ($\psi$).  It is
convenient to adopt the inclination $\alpha=20^{o}$ of Gil $\&$
Krawczyk (1997) to illustrate this point. These authors also suggest
that the polar cap emission height at 1.4MHz is $16\%$ of the light
cylinder radius. Hence the clearly-defined emission peak may be ahead
of the "true" pole by $10^{o}$ of the observed rotation phase. This
means that the notches appear $60^{o}\approx{1.0}radian$ after the
pole (or $+20^{o}$ of dipole angle after applying the factor
$\sin{20^{o}}\approx{\frac{1}{3}}$). Hence we may replace +0.5 with
-1.0 in equation (42), so that
\eq
\lab{44}
\frac{\psi}{\sin{\alpha}}=\frac{r_{*}}{r_{lc}}+1.0
\en
and we seek solutions of (40), (43) and (44) with $s<1$. No physical
solutions exist near the last closed fieldline (s=1), nor near the
polar axis (s=0), but aberration and retardation of emission on a
narrow band of fieldlines around s=0.64 can generate the observed
notches. One solution, for example, gives the absorbing site at a
distance of 1.33 light cylinder radii (400km), and the midpoint of the
absorbed emission at $1.25r_{lc}$ (370km) making an angle of
$\psi=+45^{o}$ with the dipole axis. This can just be made compatible
with the observed $3.3^{o}$ separation of the notches, which implies a
physical separation of only 20km and an absorbing site of 5km. The
aberration factor $k$ is just below unity (+0.88 from (29)),
indicating the narrowness of the parameter range within which a
physical solution can be found. Even smaller angles of $\alpha$
extends the range and raises the possible heights at which the
emission can occur.

The above "solution" clearly depends greatly on the assumption of
precise dipolar fieldlines far above the spinning star, and on our
rough generalisation of the orthogonal case to low values of
$\alpha$. Nonetheless it illustrates the point that the widened
projection of a weakly inclined pulsar's open fieldlines can permit
double notch features on the trailing edge of the profile.  Even
though the radio emission from a millisecond pulsar must come from
regions which are a significant proportion of the light cylinder
radius, the model here underlines that its weaker extended emission
may be coming from significantly higher altitudes and may well be
linked to the high-energy emission.

\section{Conclusions}
The principal conclusions of this paper can be summarised as follows:
\subsection{Geometric}
(1) Through the combined effects of aberration and retardation a
single corotating absorber can cause a system of rigidly rotating and
unidirectionally emitting sources to exhibit a double notch feature to
an external observer. This effect occurs whether the sources emit
radially or at a moderate angle to the radial direction.

(2) The absorbed emission comes from a height approximately half that
of the absorbing region. The model predicts that the width and
appearance of the double notches will be frequency-dependent but their
centroid invariant.

(3) The emitting region in the magnetosphere within which the double
notch is found must be elongated and thin relative to the scale of the
absorbing region, and be quasi-radial in structure.

\subsection{Physical}
(4) The presence of double notches in three very different nearby
pulsars demonstrates that old pulsars are capable of emitting at radio
frequencies high in their magnetospheres. There is evidence in all
three that the emission is on open fieldlines close to the dipole
axis, and may thus possibly be linked to conal or core behaviour at
the polar caps (Gil 1985). This interactive feature could be tested in
future observations.

(5) Future high-energy observations of the magnetospheres of these
pulsars may correlate with intensity-variations of their weak radio
emission. The interpretations given here suggest that a pulsar's polar
cap and its outer magnetosphere interact, and this may guide future
theoretical models (eg Harding $\&$ Muslimov 2003, Rankin $\&$ Wright
2003, Wright 2003).

(6)The emission in which notches are found is characteristically
weak, highly linearly polarised and extended over long stretches of
pulse longitude, suggesting that modelling of a $\it{relativistic}$
single vector sweep might fix the inclination of these pulsars (eg
Dyks et al 2003b).

(7) Some pulsars - maybe all - have absorbing sites located near the
light cylinder no bigger than $5\%$ of a light cylinder radius,
possibly created by a relatively high density of particles achieving
high Lorentz factors at the light cylinder and causing cyclotron
absorption (Luo $\&$ Melrose 2001, Fussell et al 2003).
 
\section{Acknowledgments}
The author thanks the University of Vermont for support from NSF grant
AST99-87654, and Joanna Rankin and Maura McLaughlin for access to
their observations and comments on the manuscript.  He is also
grateful to Drs A.Harding and J.Dyks for useful discussions, to the
referee for perceptive comments, and to the Astronomy Centre at the
University of Sussex for the award of a Visiting Research Fellowship.

\section*{References}
Arons, J. 1979, Space Sci Review, 24, 437\\
Blaskiewicz, M., Cordes, J.M., Wasserman, I., 1991, ApJ, 370, 643\\
Chen, K., Ruderman, M.A.,Zhu, T. 1998, ApJ, 493, 397\\ 
De Luca, A., Mignani, R.P., Caraveo, P.A., 2003, "Radio Pulsars" ASP 
Conference Series, eds M.Bailes, D.J.Nice, S.E.Thorsett, 302, 359\\
Donovan, J., Lommen, A., Harding, A.K., Strickman, M., Gwinn, C., 
Dodson, R., Moffet, D., McCulloch, P., 2003, IAU Symposium 218, Sydney.\\
Dyks, J., Rudak, B. 2003, ApJ, 598, 1201\\
Dyks, J., Rudak, B., Harding, A.K. 2003a, ApJ, submitted (astro-ph/0307251)\\
Dyks, J., Harding, A.K., Rudak, B. 2003b, ApJ, submitted (astro-ph/0401255)\\
Everett, J.E., Weisberg, J.M, 2001, ApJ, 553, 341\\
Fussell, D., Luo, Q., Melrose, D. 2003, MNRAS, 343, 1248\\
Gil, J., 1983, A\&A, 127, 267\\
Gil, J., 1985, ApJ, 299, 155\\
Gil, J., Krawczyk, A., 1997, MNRAS, 285, 561\\
Hankins, T.H., Cordes, J.M., 1981, ApJ, 249, 241\\
Hankins, T.H., Fowler, L.A., 1986, ApJ, 304, 256\\
Harding, A.K., Muslimov A.G., 2003, ApJ, 588,430\\
Luo, Q., Melrose, D.B., 2001, MNRAS, 325, 187\\
Lyne, A.G., Manchester, R.N., 1988, MNRAS, 234, 477\\
McLaughlin, M.A., Rankin, J.M., 2003 (MR), MNRAS, submitted\\
Mignani, R.P., De Luca, A., Caraveo, P.A., Becker, W.,  2002, ApJ, 580, L147\\
Narayan, R., Vivekanand, M., 1982, A\&A, 113, L3\\
Navarro, J., Manchester, R.N., Sandhu, J.S., Kulkarni, S.R., Bailes, M., 1997, ApJ, 486, 1019\\
Nowakowski, L.A., Bhat, N.D.R., Lorimer, D.R., 2003, Arecibo Observatory Newsletter No.36\\
Perry, T.E., Lyne, A.G., 1985, MNRAS, 212, 489.\\
Radhakrishnan, V., Cooke, D.J., 1969, ApJ. Lett, 3, 225\\
Rankin, J.M., 1993a, ApJ., 405, 285\\
Rankin, J.M., 1993b, ApJ.Suppl, 85, 145\\
Rankin, J.M., Rathnasree, N., 1997, J. Astrophys.Astr., 18, 91\\
Rankin, J.M., Wright, G.A.E., 2003, A\&A Reviews, 12, 1\\
Shearer, A., Stappers, B., O'Connor, P., Golden, A., Strom, R., Redfern, M., Ryan, O. 2003, Science, 301, 493\\
von Hoensbroech, A., Xilouris, K.M., 1997, A\&A, 324, 981\\
Wang, F.Y.-H, Halpern, J.P., 1997, ApJ Lett, 482, L159\\
Wright, G.A.E., 2003, MNRAS, 344, 1041\\
Zavlin, V.E., Pavlov, G.G., Sanwal, D., Manchester, R.N., Truemper, J.,
Halpern, J.P., Becker, W., 2002, ApJ, 569, 894\\
Zharikov, S., Mennikent, R., Shibanov, Yu., Koptsevic, A., Tovmassian, G., Komarova, V., 2003, 
in "Proc of International Workshop", Marsala, Italy 
Eds G.Cusumano, E.Massaro, T.Mineo , Rome: Aracne Editrice\\

\end{document}